\documentclass[
showpacs,preprintnumbers,amsmath,amssymb,prd]{revtex4}
\usepackage{graphicx}
\usepackage{rotating}
\usepackage{color}

\usepackage{amsbsy}
\usepackage{amsmath}
\usepackage{amssymb}
\usepackage{subfig}
\usepackage{wasysym}
\usepackage{booktabs}
\usepackage{rotating}

\newcommand{\be}{\begin{equation}}
\newcommand{\ee}{\end{equation}}
\newcommand{\ba}{\begin{eqnarray}}
\newcommand{\ea}{\end{eqnarray}}

\usepackage[latin1]{inputenc}
\usepackage{exscale}
\usepackage{bm}
\begin{document}
\title{A ``quasi-complete'' mechanical model for a double torsion pendulum}
\author 
{Fabrizio De Marchi$^1$
Giuseppe Pucacco$^{1}$, Massimo Bassan$^{1} \ast$,\\
Rosario De Rosa$^{4,5}$,  Luciano Di Fiore$^4$, Fabio Garufi $^{4,5}$, Aniello Grado$^{4,6}$\\
Lorenzo Marconi$^{2}$, Ruggero Stanga$^{2}$,Francesco Stolzi$^{2}$, Massimo Visco$^{3}$
}

\affiliation{$^1$ Dipartimento di Fisica, Universit\`a di Roma ``Tor Vergata'' and INFN, Sezione di Roma2}

\affiliation{$^2$ Dipartimento di Fisica ed Astronomia, Universit\`a degli Studi di Firenze, and INFN, sezione di Firenze}

\affiliation{$^3$ Istituto di Astrofisica e Planetologia Spaziali - INAF, Roma}

\affiliation{$^4$ INFN Sezione di Napoli -  Napoli, Italy}

\affiliation{$^5$ Dipartimento di
Fisica, Universit\`a degli Studi di Napoli "Federico II"}
\affiliation{$^6$ Osservatorio Astronomico di Capodimonte - INAF, Napoli}
\email[Corresponding author ]{bassan@roma2.infn.it}

\begin{abstract}

We present a dynamical model for the double torsion pendulum {nicknamed} `PETER',  where one torsion pendulum hangs in cascade, but off-axis, from the other.  
The dynamics of interest in these devices lies around the torsional resonance, that is at very low frequencies (mHz). However, we find that, in order to properly describe the forced motion of the pendulums, also other modes  must be considered, namely
swinging and bouncing oscillations of the two suspended masses, that {resonate} at higher frequencies (Hz).

Although the system has obviously 6+6 Degrees of Freedom,
we find that 8 are sufficient for an accurate description of the observed motion.
This model  produces 
reliable estimates of the response to generic external disturbances and actuating forces or torques. 
In particular, we compute the effect of seismic floor motion (`tilt' noise) on the low frequency part of the signal spectra 
and show that it properly accounts for most of the measured low frequency noise.

\end{abstract}
\maketitle




\section{Introduction}\label{intro}

Free Fall, i.e. motion in absence of external forces,  is a key ingredient of many present \cite{GRACE} and future \cite{GRACE2, LISA} space missions. It is particularly important for proposed LISA-like missions, that aim to detect gravitational waves by measuring the change of distance between two Test Masses in geodesic motion. Indeed the precursor LISA-Pathfinder \citep{LTP1,LTP} will soon  be launched to  demonstrate many of the key technologies that LISA is based on, including the  effective realization of a free falling Test Mass (TM), reducing the residual disturbances to the level of $10 fN /\sqrt{Hz}$ at $1 mHz$.
Such demanding requirements need ground based facilities for
preliminary free fall tests. This has prompted an extraordinary
effort into reducing the effects of gravity and other local forces.
The ideal tool for these tests is, traditionally,  the torsion pendulum, where gravity is balanced by a supporting fibre and rotational motion is virtually uncostrained  around it.  Indeed, the group at the University of Trento has extensively tested residual forces on two different apparata:  a torsion pendulum where the TM moves freely in rotation \cite{1masse} and one where the ``soft" motion is approximatively a linear translation \cite{4masse}.
The opportunity to experiment with a TM being in quasi free fall over more than one degree of freedom (DoF) is quite desirable, as it better represents geodesic behavior and it allows investigation on possible spurious couplings between motion on different DoFs. 	
This is crucial for the drag-free operation of the Test Masses \cite{dragfree} on  LISA-Pathfinder (and of any of the possible realizations of the LISA concept): in such space mission,  the TMs will move freely in the sensitive direction, but will be under feedback control on the remaining 5 directions.  An instrument with two $soft$ DoF, like a  double torsion pendulum can observe the free motion of one DoF when the other is actively controlled and can provide useful information on the amount of cross-talk that the control electronics might feed on the ``free" mode.


To this purpose, we have developed an instrument where one torsion pendulum hangs, off-axis, from another. This results in the TM being almost free (or having $soft$ restoring forces) both on a rotational and on a translational degree of freedom. The dynamics of this system is quite more complex than that of a single DoF torsion pendulum and  its description requires a careful modelling of all its  ($2 \times 6$, in principle) DoF, because some of the external disturbances (tilt motion, for example) couple to torsion via higher frequency modes, that need be accounted for in the model.
We present an analytical mechanical model of the double torsion pendulum, taking as reference the instrument built and operated in the INFN lab in Firenze (I), nicknamed  PETER (PEndolo Translazionale E Rotazionale, namely {\it Translational and Rotational PEndulum})\cite{Marconi, PETER2}.

In this instrument, a cubic TM (see the left panel in figure~\ref{angoli_cinesi}) is suspended through a double stage system with two torsion fibres in cascade: the lower fibre allows this mass to be almost free in rotation around its symmetry vertical axis. The fibre hangs from the tip of one arm of a crossbar that is, in turn, suspended with an upper fibre to the fixed frame, represented by the vacuum enclosure. The torsion of this upper fibre allows almost free motion of the suspended system, including the test mass, along an arc of a $30 \, \textrm{cm}$ diameter  circumference: for small torsion angles,  this can be considered for all practical purposes a translational motion.  Dummy loads hang from the three other arms of the crossbar. The TM is enclosed in the \emph{Gravitational Reference Sensor} (GRS)  {\cite{GRS}, a hollow metal box padded with electrodes that permit to monitor the
motion of the TM along all its translational and rotational DoF. 
 
{The TM is a
hollow Al  cube with a $46~\textrm{mm}$ edge plus a shaft (81.5 mm long) that connects to the fibre and provides electical insulation. 
The cubic TM (but not its shaft), the GRS and the readout electronics, well match}
 the set-up of the LISA-Pathfinder flight model geometry. 
 The apparatus is also equipped with additonal readouts: an autocollimator and
 an Optical Read Out (ORO) system \cite{oro} that provide independent measurements of the test mass along the $2$ soft DoF. 
 The sensitivity  goal for this apparatus, when limited by  the mechanical thermal noise  and the readout noise,  is better than $
 10^{-13} \textrm{m\, s}^{-2}/\textrm{Hz}^{1/2}$ around $1\textrm{mHz}$ (on each DoF), namely, only 1 order of magnitude worse than the LISA-Pathfinder goal along the sensitivity axis.   A more detailed description of the apparatus can be found in 
 \cite{Stanga}.

Aim of this paper is to develop an analytical model of the whole system by which its main features can be evaluated and compared with preliminary data runs.
{In the title the quotation marks around ``quasi-complete'' stand to point out that the model goes as far as possible in the comparison with the actual experiment. Although the operation range is in the mHz band around the torsional resonances, the model includes the treatment of the swinging pendulum and bouncing resonances, that take place at much higher frequencies ($\sim$Hz).

Such a detailed description may appear superfluous: however, we have carried out the analysis of the mid-high frequency range for an overall validation of the model and with the purpose of analysing possible sources of external disturbances. In particular, in view of the complex structure of PETER, the tilt noise at low frequencies can be properly described only if the coupled swinging pendulum motions are taken into account.}

 
 The plan of the paper is as follows:  in sect. \ref{mmodel} we lay out the mechanical model, with special attention to defining reference frames that are suitable to describe motion of the two payloads (the crossbar and the Test Mass) in the appropriate limit of small oscillations. The dynamical properties of the double pendulum, i.e. angular velocities and moments of inertia are then introduced and, from those, the Lagrangian of the system is derived.  Sect. \ref{sect_nm} analyizes the free motion of the double pendulum and derives the normal modes and their resonant frequencies, while in sect. \ref{sect_gf}, introducing the generalized forces on the system, we discuss how the influence of external disturbances on the system can affect the pendulum output.


\section{Eight degrees of freedom mechanical model}\label{mmodel}

In principle, in order to describe a double pendulum of the  PETER kind, with two ``payloads" suspended in cascade through two torsion fibres, we should consider 6 + 6 DoF, to allow each payload to translate and rotate in every possible way.  This, in the constrained motion of a pendulum, maps into describing the torsion, 2 pendulum motions (in 2 orthogonal directions), bouncing of the fibre length and 2 rocking (rotations around a horizontal axis passing through the suspension point) motions.
However, in all data gathered so far, the rocking modes, expected at the frequencies of 0.55Hz and 3.4Hz, were not detected: if they exist, their amplitude is well below the noise level.
Therefore, in our analysis, we neglected the rocking motion of both the crossbar and the TM, assuming that they rigidly move with the respective fibres that, therefore do not bend at the suspension point. While their inclusion would lead to more cumbersome expressions for the equations of motion, as a matter of principle they can be included without problems.

The `quasi-complete' model for PETER describes then the torsional and pendulum oscillations of the  crossbar and the TM, including also their `bouncing' motions. Therefore we deal with a 4+4 = 8 DoF model.

%

\subsection{Reference frames and configuration variables}
\begin{figure}[h!]
\centering
\includegraphics[width=0.49\columnwidth]{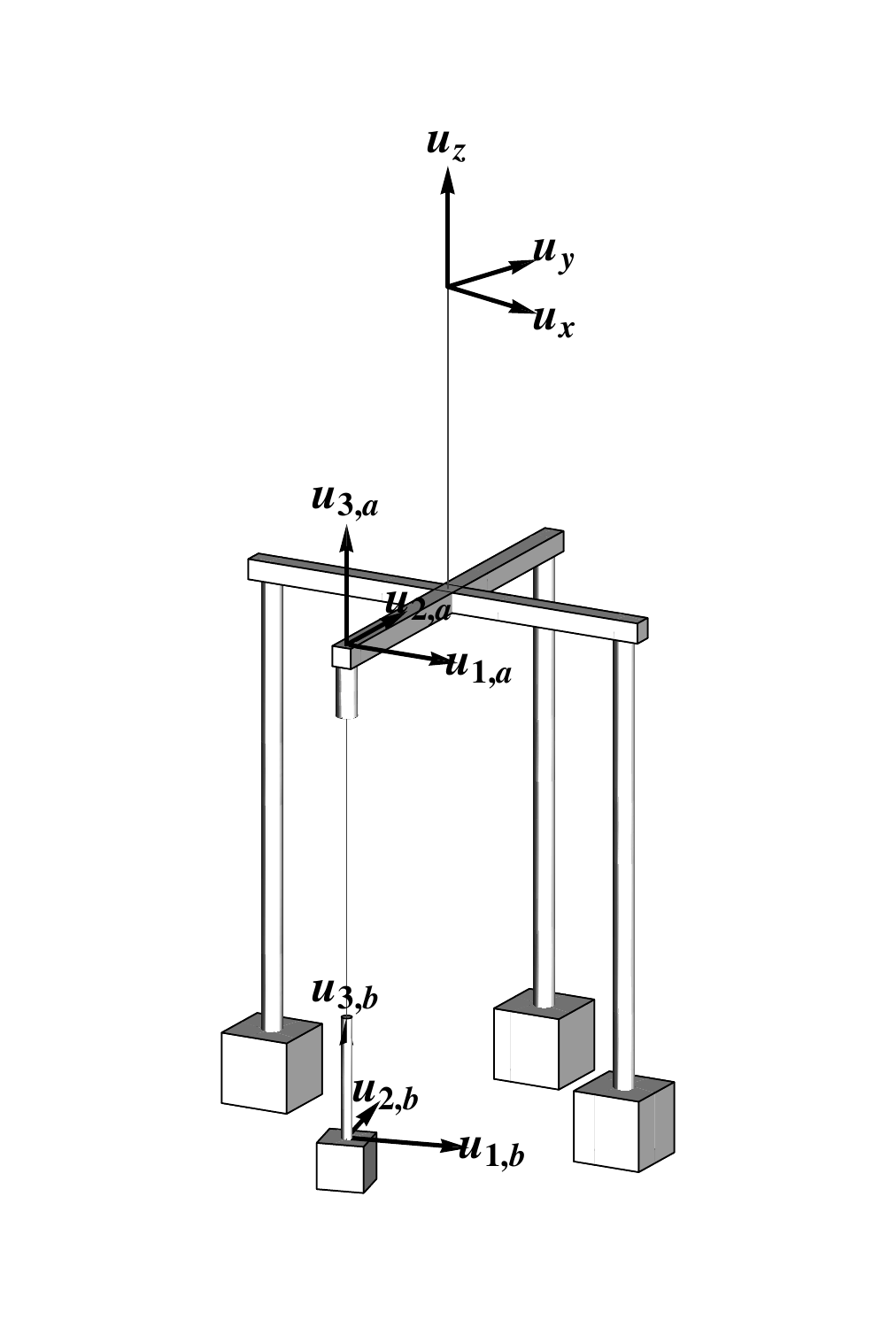}\vspace{1cm}
\includegraphics[width=0.36\columnwidth]{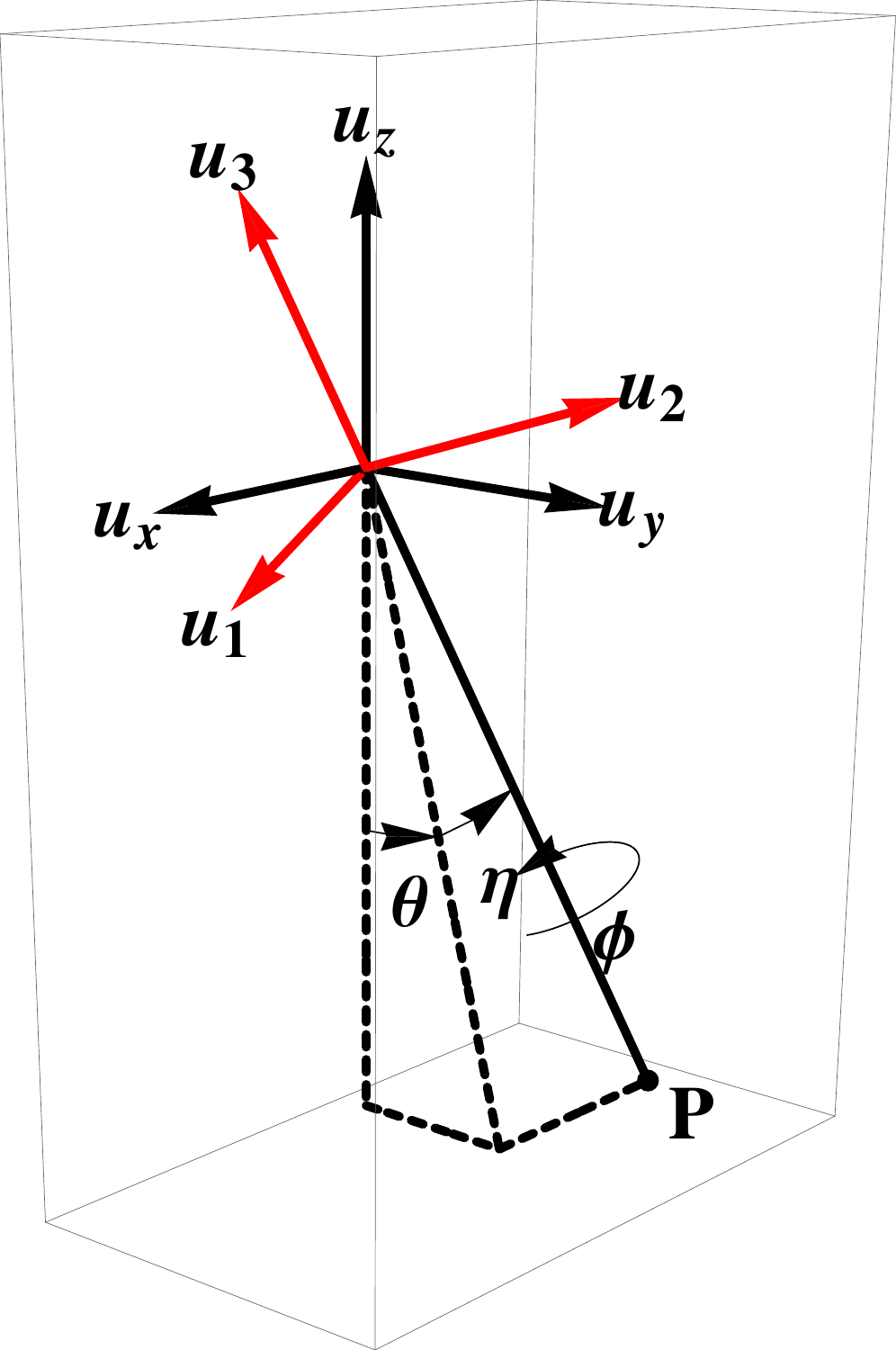}
\caption{Left panel. Schematic representation of PETER with the three triads of unit vectors. Right panel. The coordinate systems described in the text: $\bm{u}_x,\bm{u}_y,\bm{u}_z$ represent the inertial frame;   $P$ is the position of the centre of mass of the crossbar, $\bm{u}_{1},\bm{u}_{2}$ and $\bm{u}_{3}$ is one of the moving coordinate systems; $\theta$, $\eta$ and $\varphi$ the three angles to identify the orientation of the frames.}
\label{angoli_cinesi}
\end{figure}

We need to describe the position and rotation of two rigid bodies (the crossbar and the TM) with respect to an inertial frame. However, we find it useful to introduce also two moving (non-inertial) frames and exploit the Lagrangian formalism to combine generalized coordinates for a simpler description of the dynamics. {The two additional frames are comoving with the two load masses and will be labeled with the ``a" (above) and ``b" (below) subscripts.}  In analogy with the treatment of the 1 DoF case \cite{1DoF}, we have made a choice of the angles suitable to describe small oscillations around the stable equilibrium of the whole system \cite{cinesi1,cinesi2}. 

We set the origin of the inertial frame in the suspension point of the upper fibre. The orientation of this frame is as follows (see the left panel of fig.\ref{angoli_cinesi}): $\bm{u}_z$ along the local gravity acceleration
$-\bm{g}$}, $\bm{u}_x$ and $\bm{u}_y$ along the crossbar arms at the equilibrium position (when no torsion is present in the upper fibre). 
The first  non inertial frame, comoving with the crossbar (the ``$a$" frame, $\bm{u}_{1,a},\bm{u}_{2,a},\bm{u}_{3,a}$) has its origin at the end of the crossbar arm from which the lower fibre is suspended; $\bm{u}_{2,a}$ is parallel to this arm and directed towards the centre, $\bm{u}_{1,a}$ is parallel to the other arm. Since we assume no rocking motion, $\bm{u}_{3,a}$ is at all times parallel to the upper fibre direction.

The second moving frame (the ``$b$" frame, whose origin coincides with  the center of mass of the TM)  describes the orientation of the TM with respect to the  ``$a$" frame (i.e. to the crossbar).
The unit vectors $\bm{u}_{1,b}, \bm{u}_{2,b},\bm{u}_{3,b}$ are directed along the principal axes of inertia of the TM, and therefore $\bm{u}_{3,b}$ is parallel to the lower fibre direction. 
In order to bring the ``$a$"  frame 
{with its axes parallel to those of} the inertial frame, we introduce
the following sets of rotations (see the right panel of fig.\ref{angoli_cinesi}):  first, by an angle $\theta_a$ around  the $x$-axis; then by an angle $\eta_a$  around  the {new $y'$}-axis; finally, by an angle $\varphi_a$ around the  new $z'$-axis, now the same as ${\bf u_3}$. 
The resulting orthogonal rotation matrix $\bm R_a$ is\\

%
\begin{equation}
\hspace{-.1cm}
\resizebox{.90\textwidth}{!}{
$\bm{R}_a=
\begin{pmatrix}
\cos \eta_a  \cos \varphi_a &\sin \eta_a \sin \theta_a  \cos \varphi_a +\cos \theta_a \sin \varphi_a&\sin \theta_a \sin \varphi_a-\sin \eta_a  \cos \theta_a \cos \varphi_a\\
-\cos \eta_a \sin \varphi_a&\cos \theta_a \cos \varphi_a -\sin \eta_a \sin \theta_a  \sin \varphi_a&\sin \eta_a \cos \theta_a \sin \varphi_a +\sin \theta_a \cos \varphi_a\\
\sin  \eta_a&-\cos \eta_a \sin \theta_a&\cos \eta_a \cos \theta_a
\label{matriceR}
\end{pmatrix}
$}
\end{equation}


Analogously, a second rotation by the angles $\theta_b,~ \eta_b,~\varphi_b$, defining a matrix $\bm R_b$, is used for the transformation $a \rightarrow b$ 
{between} the two moving frames. 

We now introduce an additional convention: the superscripts $'$ and $''$  indicate that the components of a certain vector are expressed in the $a$ or $b$ frame, respectively. An absence of superscripts indicates that the vector is defined in the inertial frame. 
In the inertial frame, the unit vectors $\bm u_{1,a},\bm  u_{2,a}, \bm u_{3,a}$  are:
\[
\bm u_{1,a}=\bm R_a^T  (1,0,0)^T; \hspace{0.3cm}\bm u_{2,a}=\bm R_a^T  (0,1,0)^T; 
 \hspace{0.3cm}  \bm u_{3,a}=\bm R_a^T  (0,0,1)^T.
\]
The unit vectors at rest with the TM are given,in the ``$a$" frame,  by
\[
\bm u'_{1,b}=\bm R_b^T  (1,0,0)^T,\hspace{0.2cm}\bm u'_{2,b}=\bm R_b^T  (0,1,0)^T,\hspace{0.2cm}\bm u'_{3,b}= \bm R_b^T (0,0,1)^T
\]
while the same vectors are expressed in the inertial frame as:
\[
\bm u_{i,b}=R_a^T  \bm u'_{i,b} \hspace{1cm}\mbox{with}\hspace{0.5cm} i=1,2,3.
\]

\begin{figure}[h!]
\centering
\includegraphics[width=0.39\columnwidth]
{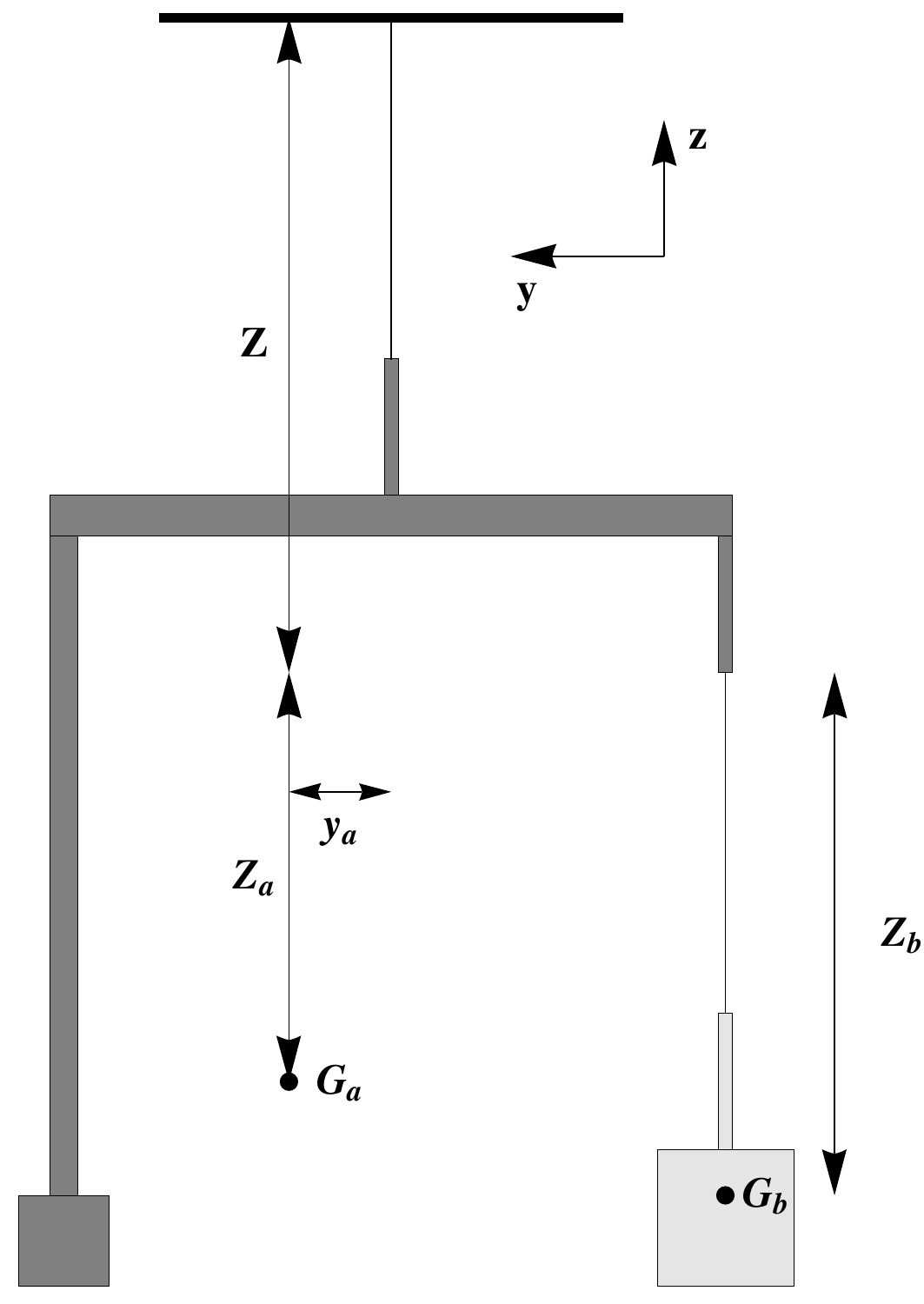}
\caption{Schematic representation of the heights of the centers of mass of the double pendulum. $G_a$ and $G_b$ are the centers of mass of crossbar+counterweights and of the TM, respectively.}
\label{schema}
\end{figure}

We now need to express the position of the center of mass, $r_{g,a}$, of the crossbar system and $r_{g,b}$ of the TM in the inertial frame (see fig. \ref{schema}).

We define ${\cal Z}$ as the component along $\bm u_{3,a}$ of the distance between
the fibres suspension points,
 ${\cal Z}_a$ as the distance, again along $\bm u_{3,a}$,  between the the lower fibre suspension point and the center of mass of the crossbar system. $\delta_a(t)$ describes the elongation of the upper fibre. We get
\be
\bm{r}_{g,a}=-({\cal Z} +{\cal Z}_a+ \delta_a(t))\bm u_{3,a}+y_a \bm u_{2,a}
\ee
where $y_a$ is the horizontal displacement of the center of mass (with respect to the line of the upper fibre). 
 It can be easily computed by assuming no misalignments: in this case the center of mass of the whole system (at the equilibrium position) has null $y$-coordinate, therefore
\[
y_a=d\frac{m_b}{m_a}
\]
where $d$ is the length of a crossbar arm, $m_a$ is the total mass of the crossbar system and $m_b$ is the mass of the TM.

Finally, if ${\cal Z}_b$ is the distance between the lower fibre suspension point and the centre of mass of the TM and $\delta_b(t)$ is the elongation of the lower fibre, for the Test Mass we obtain
\be
\bm{r}_{g,b}=-({\cal Z}+ \delta_a(t))\bm u_{3,a}-d \bm u_{2,a}-({\cal Z}_b+ \delta_b(t))\bm u_{3,b}
\label{rgb}
\ee

\subsection{Angular velocities}
Using the Poisson formula for the time derivative of a unit vector, one can readily derive
the angular velocities of the crossbar system and of the TM in the fixed frame:
\[  \bm \omega_a=\frac{1}{2}\sum_{i=1}^3 \bm u_{i,a} \times \frac{d \bm u_{i,a}}{dt}, \quad\quad  
\bm \omega_b=\frac{1}{2}\sum_{i=1}^3 \bm u_{i,b} \times \frac{d \bm u_{i,b}}{dt}
\]
In the moving frames, they are respectively transformed to:
\[
\bm \omega'_a=\bm R_a\cdot \bm \omega_a, \qquad \qquad  \bm \omega''_b=\bm R_b \bm R_a\cdot \bm \omega_b.
\]

\subsection{Inertia tensors}

We denote with $\bm{I}_a$ 
the inertia matrix for the
crossbar element, containing three counterweights and their rigid connection to
the crossbar, 
and with $\bm{I}_b$ 
that for the TM element, which includes the TM as well as the shaft connecting it to fiber b.
Both inertia matrices are calculated with respect to their centers of mass and in the references frames ($ \bm u_{1,a} , \bm u_{2,a} , \bm u_{3,a} $) and ($ \bm u_{1,b} , \bm u_{2,b}, \bm u_{3,b} $), respectively. The plane $x=0$ is a symmetry plane for the crossbar system, therefore $I^a_{12}$ and $I^a_{13}$ are expected to be zero or, at least, quite smaller than the others coefficients. Instead, the inertia matrix of the TM is expected to be diagonal and with two terms equal to each other.
\[
\bm I_a=\left(
\begin{array}{ccc}
I^a_{11} & 0 & 0 \\
0 & I^a_{22}  & I^a_{23} \\
0 & I^a_{23} & I^a_{33}\\
\end{array}
\right),
\qquad \bm I_b=\left(
\begin{array}{ccc}
I^b_{11} & 0 & 0 \\
0 & I^b_{11}  & 0 \\
0 & 0 & I^b_{33}\\
\end{array}
\right) .
\label{}
\]


\subsection{Lagrangian and generalised coordinates for the system}

A rigid body with inertia tensor $I_{jk}, \; j,k=1,2,3$ and angular velocity $\omega_k$ has rotational kinetic energy

\[
K_{rot}=\frac{1}{2}\sum_{jk}I_{jk}\omega_j \omega_k = \frac{1}{2}\bm \omega \cdot \bm{I} \cdot \bm \omega.
\]
The translational kinetic energy is instead obtained from the velocity of the centre of mass:
by using the above defined coordinates, we therefore obtain:
\[
K_{transl,a}= \frac{1}{2} m_a \vert \dot{\bm{r}}_{g,a} \vert^2, \hspace{0.95cm}K_{transl,b}= \frac{1}{2} m_b \vert \dot{\bm{r}}_{g,b} \vert^2
\]
and

\[
K_{rot,a}=\frac{1}{2}\bm \omega'_a\cdot \bm{I}_a\cdot\bm \omega'_a, \qquad K_{rot,b}=\frac{1}{2}\bm \omega''_b \cdot \bm{I}_b\cdot \bm \omega''_b.
\]
The total potential energy is

\[
U=m_a g z_{g,a}+m_b g z_{g,b}+\frac{1}{2}k_a \varphi_a^2+\frac{1}{2}k_b \varphi_b^2+\frac{1}{2}\kappa_{e,a} \delta_a^2+\frac{1}{2}\kappa_{e,b} \delta_b^2,
\]
where 
$z_{g,a}$ and $z_{g,b}$ are respectively the heights of the centers of mass of crossbar and TM,  $k_a$ and $k_b$ are the torsional constants of the two fibres and $\kappa_{e,a}$ and $\kappa_{e,b}$ are the bouncing spring constants, responsible for the fibre elongation \cite{1DoF}. 
The fibres are considered here as ideal elastic wires, as discussed in sect.\ref{Seismic}. Therefore, we neglect bending contributions to the potential energy.

The total Lagrangian of the system is then
\[
 {\cal L} (\vec q, \dot {\vec q})= K_{transl,a} + K_{transl,b} + K_{rot,a} + K_{rot,b} - U, 
\]
where with $\vec q$ we denote the set of configurational coordinates $\theta_a$, $\eta_a$, $\varphi_a$, $\delta_a$,  $\theta_b$, $\eta_b$, $\varphi_b$, $\delta_b$.  This Lagrangian is only apparently decoupled in its variables:  indeed, the coupling is hidden in the definition of $r_{g,b}$ (eq.\ref{rgb}).
We now exploit the limit of motions of small amplitude, and construct a quadratic Lagrangian from which we calculate the equations of motion {(reported in Appendix)}.

In the limit of small oscillations, the observable quantities (position and orientation of the TM as measured by the GRS) are related to the Lagrangian coordinates by
\be
\begin{pmatrix}
X\\
Y\\
Z\\
\end{pmatrix}
\simeq 
\begin{pmatrix}
-\eta _a \left({\cal Z}+{\cal Z}_b+h_c\right) -\eta _b\, (h_c+{\cal Z}_b)+\varphi _a d\\
\theta _a \left({\cal Z}+{\cal Z}_b+h_c\right)+\theta _b \ (h_c+{\cal Z}_b)\\
 -\delta _a-\delta _b-\theta_a d
\end{pmatrix};\quad
\ee

\[
\begin{pmatrix}
\theta\\
\eta\\
\varphi\\
\end{pmatrix}
\simeq 
\begin{pmatrix}
\theta_a+\theta_b\\
\eta_a+\eta_b\\
\varphi_a+\varphi_b\\
\end{pmatrix}
\label{eq_coord3}
\]
where $h_c$ is 
the distance between the cube geometrical center  and the center of mass of the TM torsion member (i.e. $G_b$ in Figure \ref{schema}) , which  are offset due to the mass of the shaft connecting the TM to the torsion fibre.\\
We remark here  that $\varphi_a$ is an additional observable quantity, as it can be monitored separately, e.g. via an autocollimator measuring rotation of the crossbar. Moreover, it is a $good$ observable, because it relates in a simple, straightforward way to any linear force acting on the TM.

\section{Normal modes and free motion}\label{sect_nm}

To validate the model we now compare its predictions with real data from the PETER apparatus \cite{parigi}.
Lengths, masses and inertia moments are measured or computed with good accuracy. Springs constants are derived from measurements of resonant frequencies of each oscillator. However, when analyzing a 2 DoF system, we
observe the resonances of the coupled modes, rather than those of the physical oscillators. For this reason, we used preliminary data
when  the crossbar was clamped and 
we effectively only had the isolated ``$b$"  pendulum, with
the TM moving along four DoF (2 swinging modes, 1 torsional mode and 1 bouncing mode). We measured the resonant frequencies: 
$\nu_{torsional}=2.2$ mHz, $\nu_{bouncing}=8.82$ Hz .\\
From these measured values, we derived (see \cite{1DoF} for details) the 
torsional and bouncing spring constants  of the ``$b$" fibre, reported in table 1.
We then estimated the spring constants of  the ``$a$" pendulum, by applying 
the following scaling relations, involving the
radii ($r_a , r_b$) and lengths ($l_a,l_b$) of the fibres,:
\[
k_a=k_b \left(\frac{r_a}{r_b}\right)^4 \frac{l_b}{l_a}; \qquad \kappa_{e,a}=\kappa_{e,b} \left(\frac{r_a}{r_b}\right)^2 \frac{l_b}{l_a}
\]
The lengths of the fibres are re-defined by taking into account the static longitudinal deformation (a few mm in both cases) due to their respective loads.
The torsion constant of the  `$a$" pendulum was measured  on an independent, dedicated test apparatus, yielding a value in excellent agreement with the above determination via scaling.
All physical and geometrical parameters of the double pendulum are summarized in Table \ref{numval}; measurement errors amount to a few percent.
By using these values we compute the frequencies of the normal modes, shown in the second column of Table \ref{normalmodes}.  A least square fit, constrained within the experimental error bars, then adjusts the mechanical parameters in order to best match the observed normal mode frequencies, reported in column 3 of the same Table.
\begin{table}[thb]
\hspace*{-1cm}
\begin{tabular}{ll}
\hline
 \multicolumn{2}{l}{{\bf Inertia matrices}  (units= [kg m$^2$])} \\ 
 $I^a_{11}$ = 0.182 &  $I^b_{11}$ =$2.76\cdot 10^{-4}$ \\
 $I^a_{22}$ = 0.184 &  $I^b_{22}$ = $2.77\cdot 10^{-4}$ \\
$I^a_{33}$ = $2.38\cdot 10^{-2}$  & $I^b_{33}$ = $3.71\cdot 10^{-5}$ \\
 $I^a_{23}$ = $-8.13 \cdot 10^{-3}$  &\\ 
  $I^a_{12}$ =  $I^b_{12}$ =  $I^a_{13}$ =  $I^b_{13}$ = $I^b_{23}$ =0  \\
\hline
 \multicolumn{2}{l}{{\bf Masses}   [kg]} \\ 
 $m_a$=1.2 & $m_b$=0.11 \\
\hline
\multicolumn{2}{l}{{\bf Torsional constants}  [kg m$^2$ s$^{-2}$] }  \\ 
$k_a=1.8\cdot10^{-6}$  & $k_b=7.1\cdot 10^{-9}$\\
\hline
\multicolumn{2}{l}{{\bf Bouncing constants}  [kg s$^{-2}$]}  \\ 
 $\kappa_{e,a}=4804$ & $\kappa_{e,b}=300$  \\
\hline
\multicolumn{2}{l}{{\bf Lengths}  [cm]}  \\ 
\multicolumn{2}{l}{$d$=15; \  \ $h_c$=3.4; \ \ ${\cal Z}$=87; \ \ ${\cal Z}_b$=76; \ \ ${\cal Z}_a$=43}\\
\hline
\end{tabular}
\caption{Numerical vaues of the mechanical parameters of} the PETER double pendulum. Masses and lengths are measured, inertia moments and spring constants are derived from measured quantities. { Experimental errors are of the order of 2-3 \% for most values, and 10\% for the torsion and bouncing constants. These quantities are used as $input$ parameters for the model, yielding the eigenfrequencies listed in table \ref{normalmodes}.} 
\label{numval}
\end{table}

\begin{table}[thb]
\begin{tabular}{|c|c|c|c|}
\hline
 \multicolumn{4}{|c|}{\bf Normal modes frequencies } \\ 
 \hline
mode \# & calculated & measured & unit\\
\hline
$\nu_1$& $1.3 \pm 0.1$ & 1.331 &mHz~ \\
$\nu_2$& $2.2 \pm 0.2$ & 2.117 &mHz~ \\
$\nu_3$& $0.41\pm0.01$ & 0.406 &Hz\\ 
$\nu_4$& $0.42~\pm 0.01$ &0.4065 & Hz \\
$\nu_5$& $0.59\pm0.01$ & 0.58815& Hz \\
$\nu_6$& $0.62\pm 0.01$ & 0.6170 &Hz \\
$\nu_7$& $8.05\pm 0.6$ & 7.926 & Hz \\
$\nu_8$&$ 10.9\pm0.8$ & 10.393 & Hz\\
\hline
\end{tabular}
\caption{Calculated and measured normal modes frequencies. The error on the calculated values are obtained by a Monte Carlo variation of the input parameters within their experimental error, given in table \ref{numval}. }
\label{normalmodes}
\end{table}

The two pairs  $\nu_{3,4}$ (virtually degenerate)  and $\nu_{5,6}$ are respectively associated to the swinging motion of the crossbar+counterweights system and of the TM.  
Indeed, if each pendulum behaved as a decoupled physical pendulum, we would get 
\begin{eqnarray*}
\nu_{3,4}\approx &\sqrt{\dfrac{m_a g ({\cal Z}+{\cal Z}_a)}{I^a_{11}+m_a({\cal Z}+{\cal Z}_a)^2}}&=0.418\, {\rm Hz};\quad\mbox{ \hskip 0.4cm  and} \\ 
\nu_{5,6}\approx &\sqrt{\dfrac{m_b g {\cal Z}_b}{I^b_{11}+m_b {\cal Z}_b^2}}&=0.568\,{\rm Hz}.\\[8pt]
\end{eqnarray*}

\begin{figure}[h!]
\centering
\includegraphics[width=0.49\columnwidth]{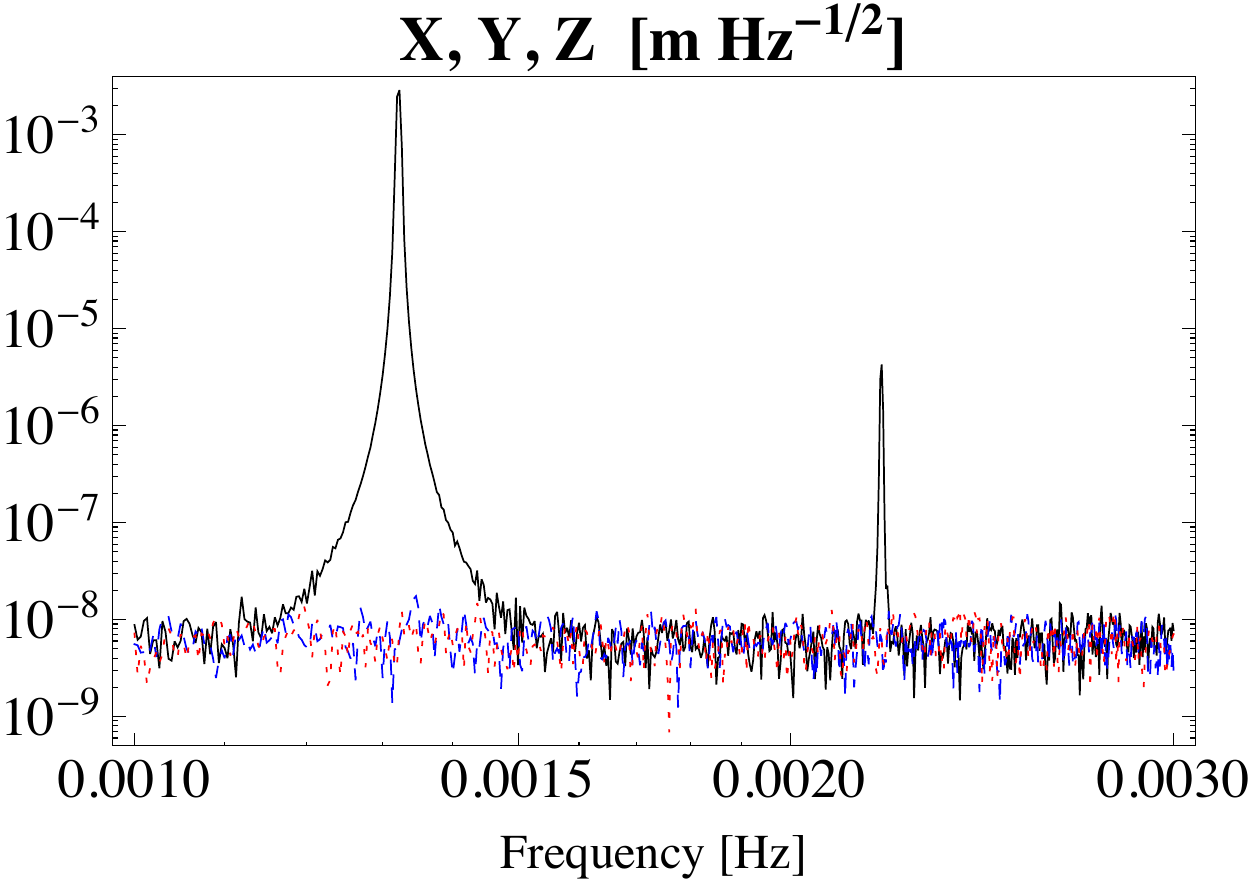}
\includegraphics[width=0.47\columnwidth]{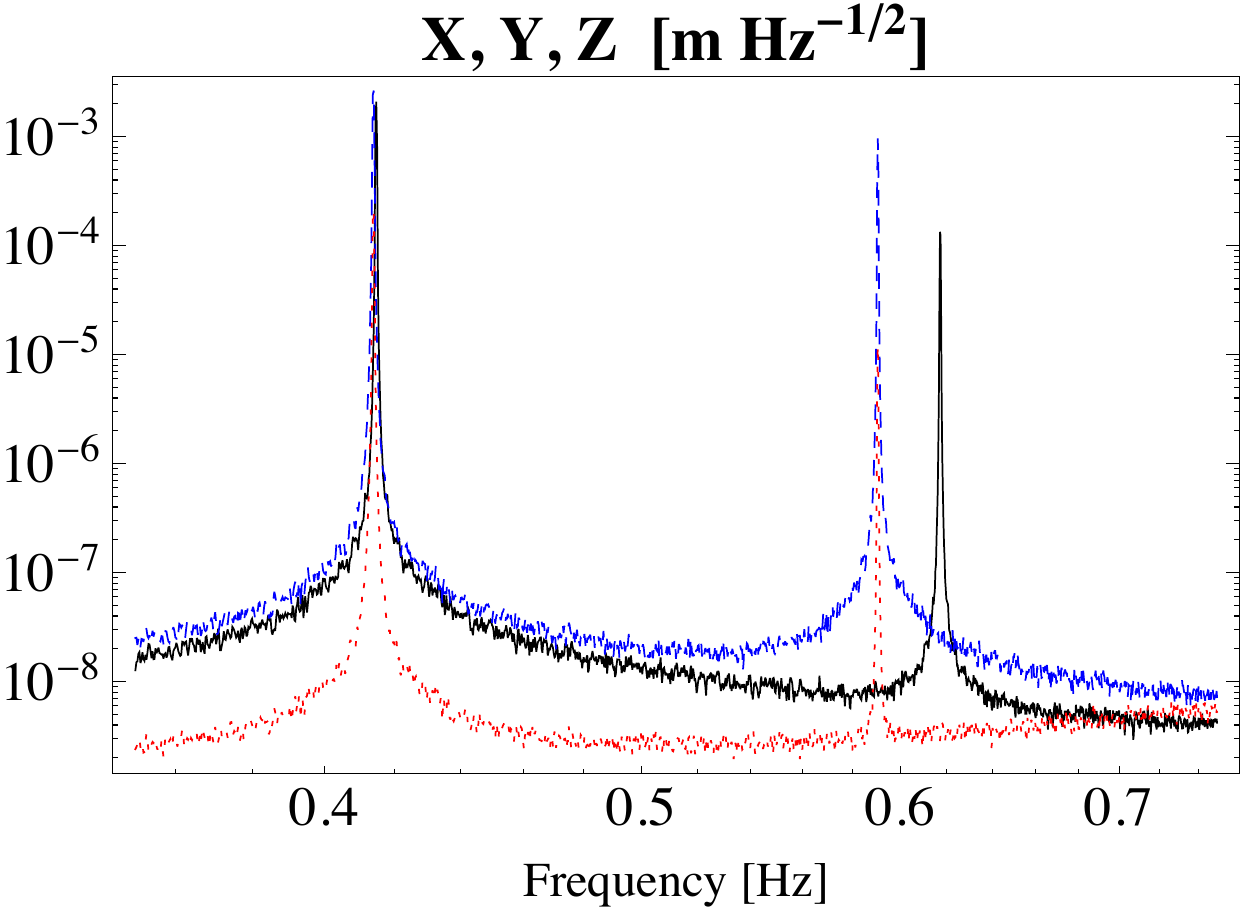}
\includegraphics[width=0.49\columnwidth]{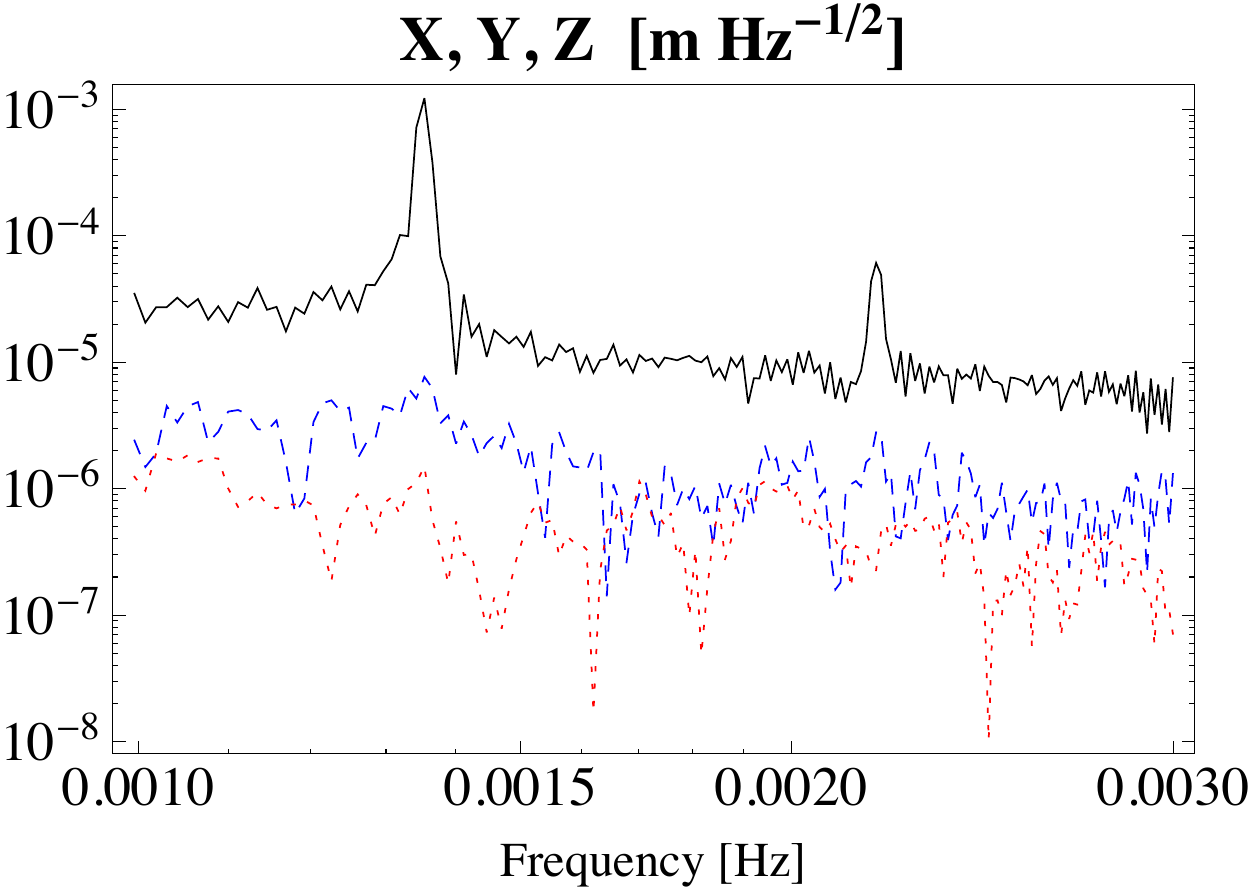}
\includegraphics[width=0.47\columnwidth]{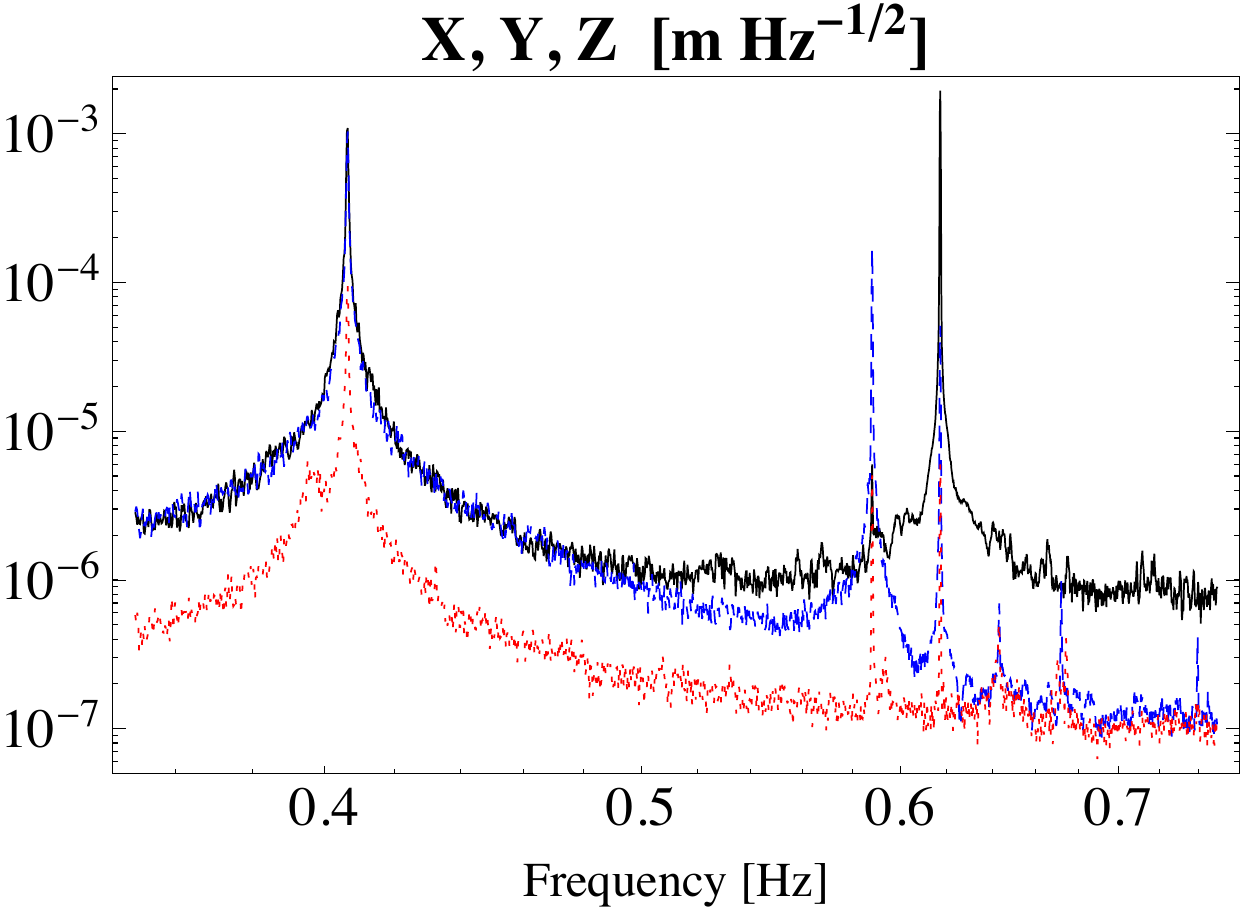}
\caption{{Amplitude spectral densities of observable displacements $X,Y,Z$.
Left:
in the frequency region around the torsional resonances.  Rigth: 
 in the intermediate frequency range, where the swinging resonances are observed.
 Upper plots show the model predictions, while the lower plots display the experimental data. 
Solid line (black online): $X$, dashed line (blue online: $Y$, dotted lines (red online): $Z$. Only the motion along $X$ is activated at the torsion eigenfrequencies.}}
\label{spettri_LF}
\end{figure}

There are two main reasons for the line splitting $\nu_5, \nu_6$, associated with the swinging of the TM: the first is that the centers of mass of both the  crossbar+counterweight system and of the TM are not aligned with the upper fibre (cfr. the case of 1 DoF, \cite{1DoF}); the second is that the inertia matrix of the crossbar+counterweight system is not diagonal. 
In other words, the ``b'' pendulum  couples differently  to the crossbar system, depending on whether  the lower pendulum swings parallel or normal to it. 
We then solved the equations of free motion (namely the  homogeneous system of equations associated to  (\ref{EM}))
with many arbitrary but  'reasonable' initial conditions. From the solutions we have obtained the observable coordinates by using Eqs. (\ref{eq_coord3}). 
The solutions were then sampled at 0.1 s, just as in the actual experiment, and from these the spectra  of the simulated data stream were computed. A white noise of $9 \, nm/\sqrt(Hz)$ was added to the model spectra, to simulate the readout noise.
Note that damping is not considered in the model; therefore, the width of the resonances is only determined by the length of the time series, that roughly corresponds to 12 hours of data.
Finally, we compared these spectra with  those of  preliminary experimental runs, as shown in fig. (\ref{spettri_LF}).  The experimental spectra have a much larger wide band noise that is partly due to tilt effects and partly under investigation.
Similar comparisons have been carried out for the rotational observables ($\varphi, \eta \, \theta$) that are measured by the GRS. \\
We note that the model well replicates  most of the features of the actual experiment, including the non trivial prediction about which resonances appear in a given observable channel. 
We verify, for example, that the swinging resonances of the lower fibre,  at 0.6 Hz in fig.\ref{spettri_LF},  are  split in a doublet ($\nu_5$ and $\nu_6$), as stated above. 
The X{(and $\eta$, not shown)}  channels only see the $\nu_6$ mode, and that is correctly predicted by the simulation.  However,  discrepancies remain: there are modes that show up in some channels in the data, while not predicted as, e. g.,  $\nu_6$ in the $Y$ and $Z$ spectra. 
We recall however that the comparison is not completely fair:  simulated data only predict the $free$ motion of the double pendulum, while the instrument is certainly driven by external disturbances. Besides, unavoidable asymmetries in the assembly of the system (e.g., an imperfect match of weigth on the four arms of the crossbar, that can lead to the crossbar laying at rest in a non-horizontal plane)  are not accounted for in the model, and can easily lead to the appearance of these modes \\
In the following section we discuss how to include driving effects, through the formalism of Lagrangian generalized forces. We apply this method,  as an example, to one of the most obvious and easily measurable of these external noise sources, i. e. floor tilt, and show how this forcing term, even if only relevant at very low frequencies, can produce excitation of higher modes.

\begin{figure}[h!]
\centering
\includegraphics[width=0.6\columnwidth]{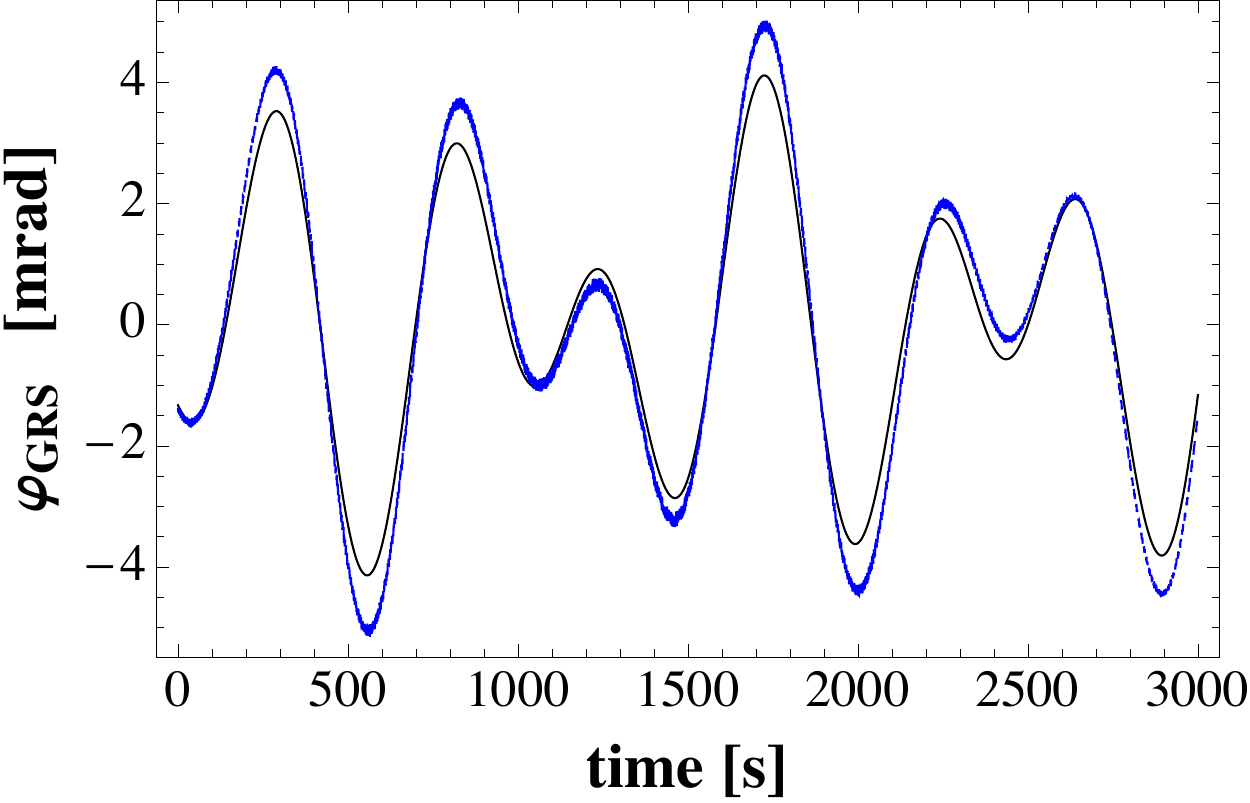}
\caption{Measured (dashed-blue) and simulated (continuous-black)  $\varphi$ evolution of a large amplitude, virtually free motion: low frequency behaviour of the model well replicates the experimental data. The slight mismatch in the amplitudes of vibration can easily be due to imperfect matching of the gain calibration coefficients ($\pm 2 \%)$ or experimental errors on the model parameters.}
\label{libero1}
\end{figure}

\begin{figure}[h!]
\centering

\includegraphics[width=0.6\columnwidth]{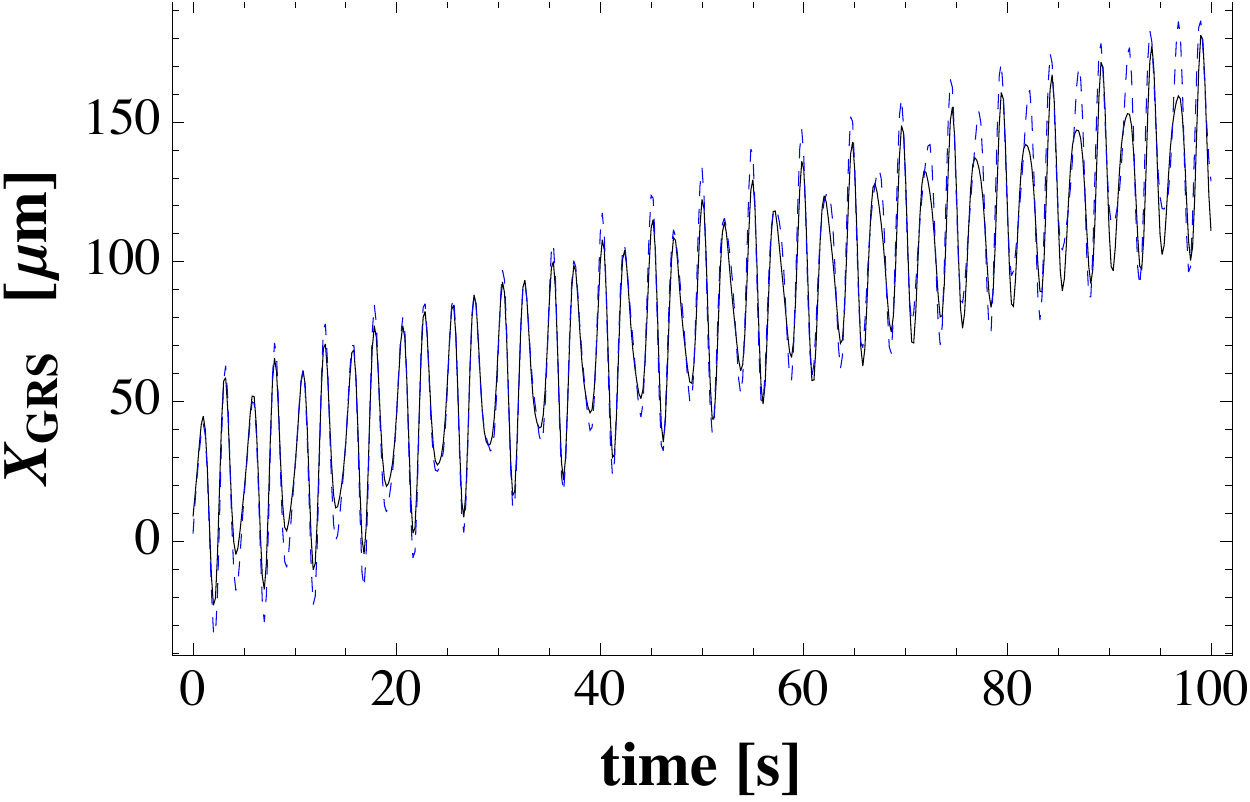}
\caption{Measured (dashed-blue) and computed (continuous-black)  $X$ evolution of  of a large amplitude, virtually free motion. A fraction of a torsion period is shown, in order to better display the motion at the higher, swinging frequencies.}
\label{libero2}
\end{figure}

In order to further validate the model, we have undertaken another comparison: the TM was moved away from equilibrium by  a large $kick$ along the X axis by the electrostatic actuation, and then released.   We obtain a good approximation of free motion, as possible disturbances are only relevant at smaller amplitudes. The same initial condition were then used for the model, and the resulting motion calculated.  Figure \ref{libero1} shows the comparison of real and simulated data for the $\varphi$ channel of the GRS: while the match of the low frequency behavior (a sum of the two normal mode oscillations at $\nu_1$ and $\nu_2$) is excellent, the measured data show a modulation at the swinging frequencies that the model does not replicate.
However, figure \ref{libero2} shows a zoom on a smaller stretch of time of the $X$ observable, showing a good agreement also for the motion at the swinging frequencies.

\section{Generalized forces}\label{sect_gf}

PETER will be not, in general, an isolated system. External disturbances such as the seismic vibration of the laboratory, parasitic electrical or magnetic fields,
mechanical dissipations etc. 
can be modelled as time-dependent forces and torques applied to the crossbar and/or the TM. 
Moreover, by changing the voltage on the electrodes one can induce on the TM a deterministic force and/or momentum (actuation). Actuations can be periodic functions, but it is also possible to induce a `viscous damping' in order to reduce the amplitude of the oscillations of the TM.

\subsection{Small external disturbances}
In the most general case, two forces ($\bm F_a$, $\bm F_b$) and  two torques ($\bm M_a$,  $\bm M_b$) with arbitrary intensity and direction must be considered. Momenta are calculated with respect to the centre of mass of the crossbar+counterweights system and the TM. In the limit of small displacements, we can perform a series expansion and keep the lowest order non-trivial terms.

In the general case of a non-isolated system of $M$ particles with $N$ DoF ($N < 3 M$ if there are constraints, in our case, simply, $N=2$), we can write the Euler-Lagrange equations in the presence of generalized forces:
\[
\frac{d}{dt}\frac{\partial {\cal L}}{\partial \dot q_h}-\frac{\partial {\cal L}}{\partial q_h}=Q_h; \quad
Q_h=\sum_{j=1}^M \bm{F}_j \cdot \frac{\partial \bm{r}_j}{\partial q_h}\quad
h=1,\dots,N
\]
where $\bm F_j$ is the force applied on the $j$-th particle. If the system consists of a set of $M'$ rigid bodies, the generalized forces are written as:
\begin{equation}
Q_h = \sum_{i=1}^{M'} \left[ \frac{\partial \bm r_{cm, i}}{\partial q_h} \cdot \bm F_{T,i}  + \frac{\partial \bm \psi_i}{\partial q_h} \cdot \bm M_i\right]
\label{qh}
\end{equation}
where the $\bm F_{T,i} $ and $\bm M_i$ are respectively the resulting external force and momentum and the ${\bm \psi}_i$ are implicitly given by
\[
\bm \omega_i= \sum_h \frac{\partial \bm \psi_i}{\partial q_h} \dot q_h.
\]
We apply eq.(\ref{qh}) to our case ($M'=2$), with two generic forces  $\bm F_a$, $\bm F_b$ and torques $\bm M_a$,  $\bm M_b$ applied on the TM and the crossbar.  We report in Appendix the explicit expressions for the $Q_h$, linearized
 in the configuration variables.

\subsection{Seismic effects} \label{Seismic}
Generalized forces can be used to perform an analysis of seismic effects on the motion of PETER.
With respect to an ideal inertial frame, the floor of the laboratory is subject to small translations and rotations ({``tilt'' noise}). 
With a simple model we calculate how the observables of PETER are affected by the motion of the ground.
 We assume, for simplicity, the vacuum chamber where PETER is housed to be a rigid body (this is certainly true at low frequencies) and we neglect the flexural stiffness of the fibres
 \footnote{
Indeed, the restoring force applied by the elasticity of the fibre is much smaller than that of gravity. They can be evaluated in terms of the fibre parameters: length $l_a$, second moment of inertia $I$,  Young modulus of W  $E$ and  tension $T$. For our W fibre 
we get: 
\[
k_{el.}=\frac{\sqrt{T E I}}{2 \,{l_a}^2}\simeq 5 ~ 10^{-4} \quad \mbox{N  m}^{-1}; \hskip 0.7cm k_{grav.}=T/l_a \simeq 21.1~ \mbox {N\,m}^{-1}
\]
}.

The tilt of the ground with respect to a``fixed" frame, is describer by an angular velocity that we express, for small angles, as $\bm \omega_t=(\dot \theta_t (t), \dot \eta_t (t), \dot \varphi_t (t) )$. We focus here on rotations, neglecting for the moment the translation of the ground,  as well as the negligible rotation on the horizontal plane, $\varphi_t (t)$.  Indeed, $(\theta_t (t), \eta_t (t))$ are directly measured by a tilt-meter and can readily provide a test for  the model.
The whole (rigid) structure rotates around its base with angular velocity $\bm \omega_t$ and, consequently, the upper fibre suspension point (that is at a distance $h$ from the ground) is affected by a linear acceleration $\ddot{\bm r}_t=h(\ddot \eta_t,-\ddot \theta_t,0)$.

We have derived the Lagrangian of the double pendulum in an inertial frame. Since gravitational potential energy depends on the elevations of Test Mass and crossbar, the coordinates $\theta_a$ and $\eta_a$ must be corrected, by adding the corresponding tilt angles, as $\theta_a+\theta_t$ and $\eta_a+\eta_t$,  while all other coordinates remain unchanged. 
After this change in the Lagrangian, 
we must add to the resulting equations of motion 
the external (apparent) forces due to acceleration of the origin and to angular velocity. 
In the resulting, well known expression for the force on a generic mass element $dm$, of coordinate {$\bm r$}, in an accelerated system:
\be
d \bm F_{app}=-d m (\ddot {\bm r_t} + \bm \omega_t \times (\bm \omega_t \times \bm r) + 2 \bm \omega_t \times \dot {\bm r}+ \dot {\bm \omega}_t \times \bm r)
\ee
we can safely neglect, for small $\bm \omega_t$ and small oscillations,  both the centrifugal 
and the Coriolis term 
that represent second order corrections.  The only relevant terms are then the first and the last one.

We now compute the total moment of the apparent forces with respect to the center of mass $\bm r_g$ of a rigid body, by integrating over its volume.  With the substitution $\bm r^*=\bm r- \bm r_{g}$, and making repeated use of the property {$\int \bm r^* dm=0$}, we obtain:

\[
\bm M_{app}\doteq\int \bm r^* \times d \bm F_{app,i}=- \int  \bm r^* \times  (\ddot{\bm r_t}+ \dot {\bm \omega}_t \times \bm r) dm.
\]
As the first term is zero, 
\[
\bm M_{app}=- \int \bm  r^* \times ( \dot {\bm \omega}_t \times \bm r_g) dm- \int  \bm r^* \times ( \dot {\bm \omega}_t \times \bm r^*) dm=-\bm I \cdot \dot {\bm \omega}_t
\]
where $\bm I$ is the inertia matrix of the rigid body.

In this way, we have derived the apparent forces and torques acting on the Test Mass and crossbar:
\[
\begin{array}{l}
\bm F_{app,a}=-m_a (\bm \ddot {\bm r_t}+\dot {\bm \omega}_t \times \bm r_{g,a}); \qquad \bm F_{app,b}=-m_b (\bm \ddot {\bm r_t}+\dot {\bm \omega}_t \times \bm r_{g,b});\\[8pt]
\bm M_{app,a} = -\bm I_a \cdot \dot {\bm \omega}_t; \qquad \qquad \qquad \quad  \bm M_{app,b} = -\bm I_b \cdot \dot {\bm \omega}_t.\\[8pt]
\end{array}
\]
Inserting these expressions into eq.(\ref{qh}), we can now explicitly compute the effect of tilt motion on the torsion pendulums observables. 

\begin{figure}[h!]
\centering
\includegraphics[width=0.49\columnwidth]{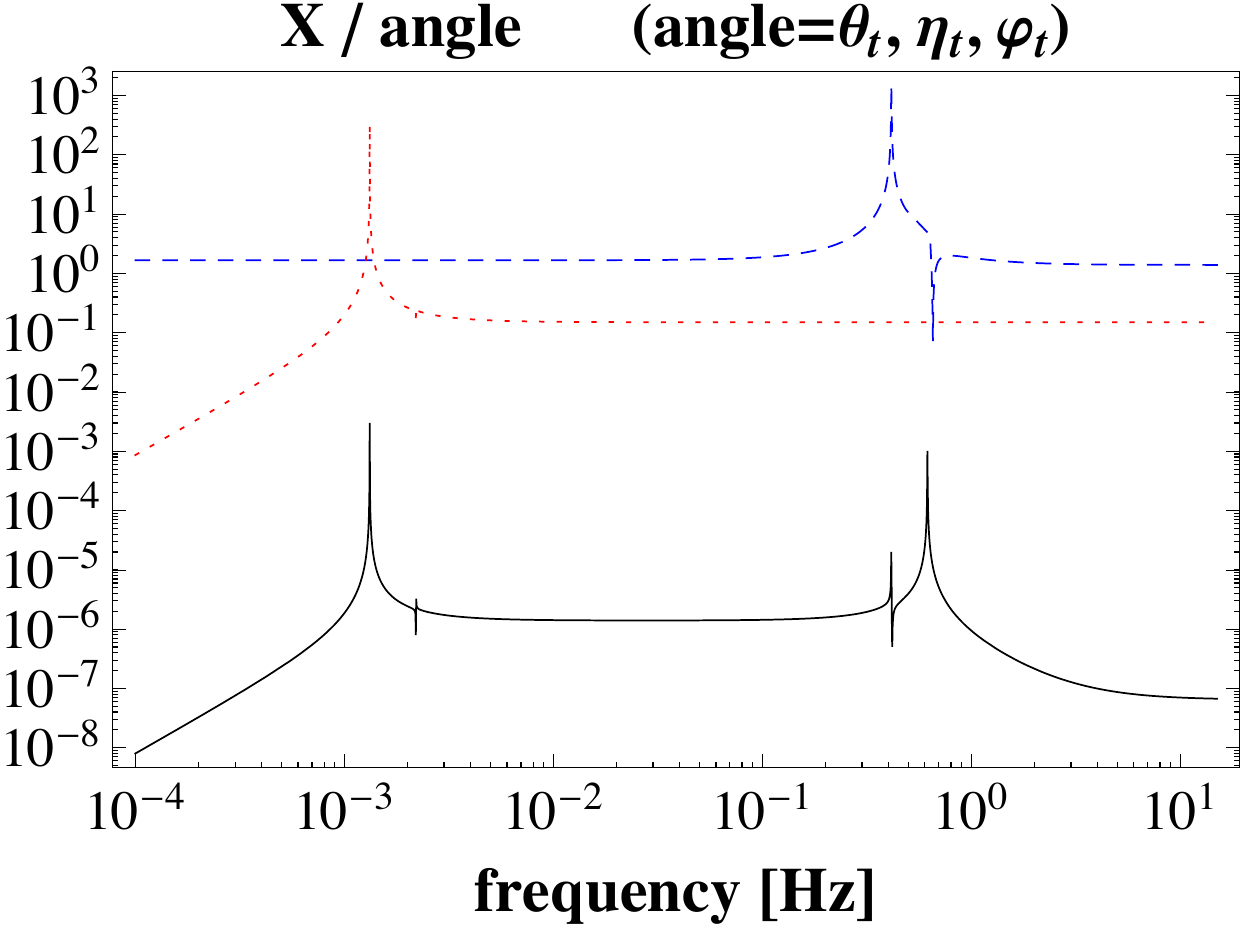}
\includegraphics[width=0.49\columnwidth]{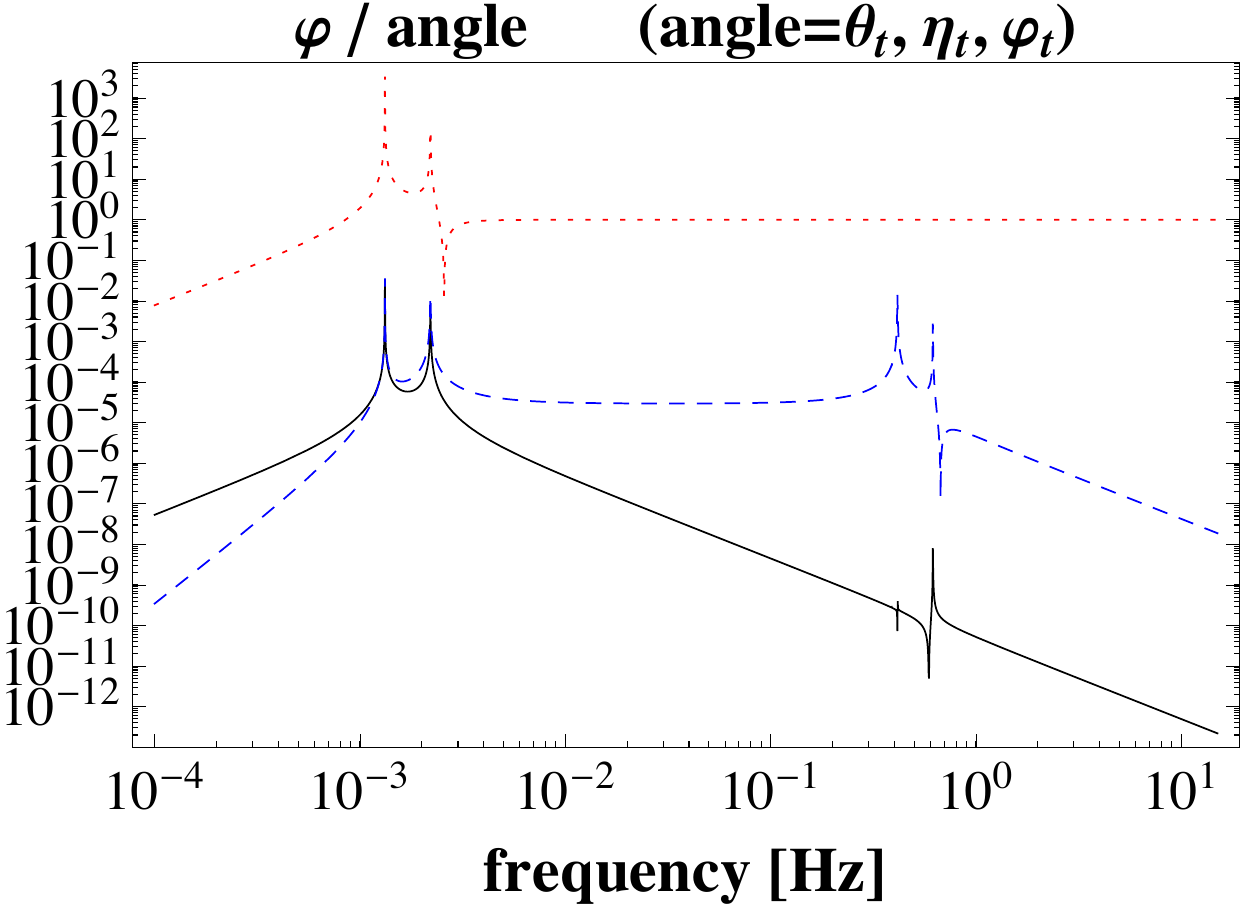}
\includegraphics[width=0.49\columnwidth]{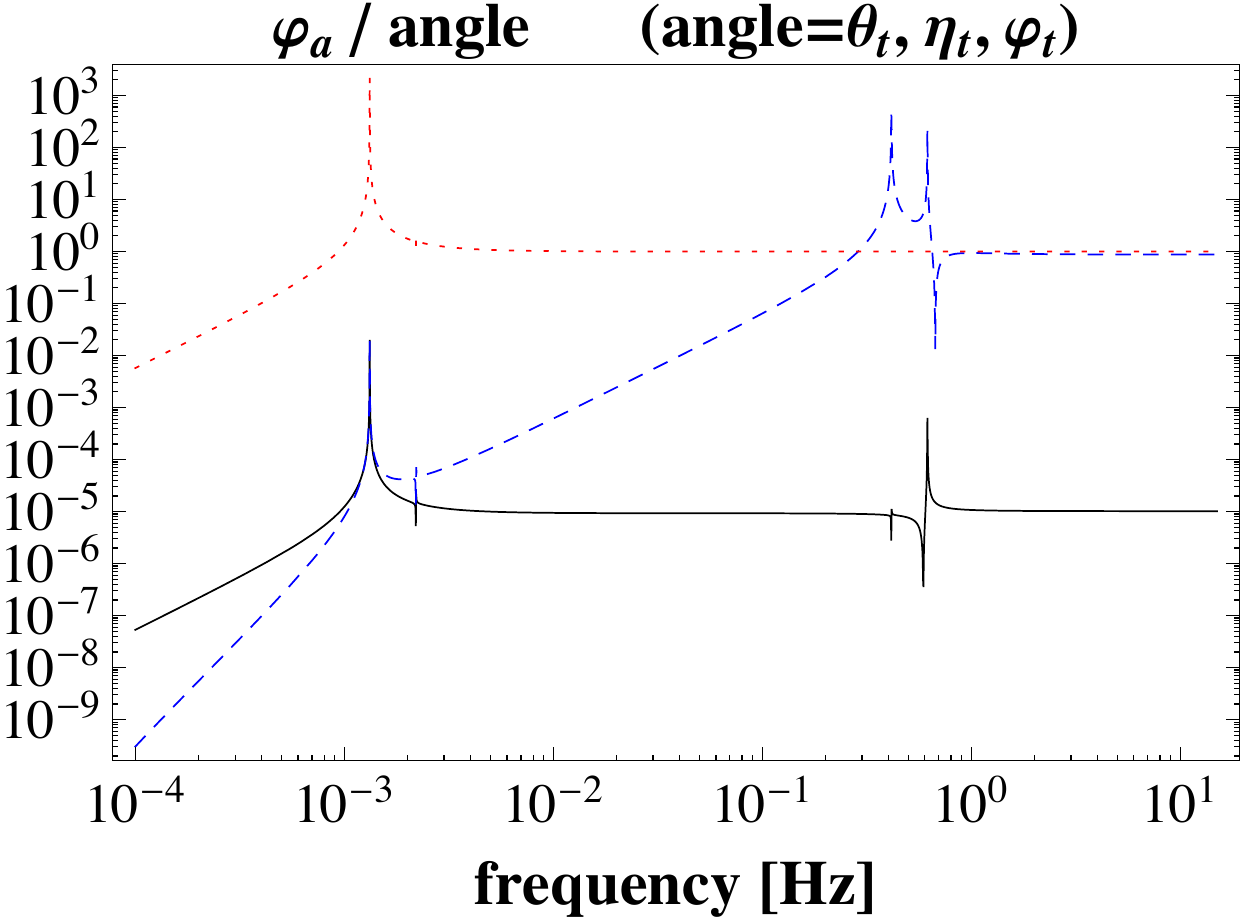}
\caption{Transfer functions for the observables $X$ (left panel), $\varphi$ (right panel) and $\varphi_a$ (bottom panel) relative to the ground tilt angles $\theta_t$ (solid lines),  $\eta_t$ (dashed), $\varphi_t$ (dotted).}
\label{tilt}
\end{figure}

We report in Figure \ref{tilt}  the transfer functions from tilt angles ($\theta_t,\eta_t,\varphi_t)$ to the $X$,  $\varphi$ and $\varphi_a$ coordinates
that monitor the ``soft" DoF and are therefore of special interest.
We observe that, when the frequency approaches zero (below the torsional resonances),
$X\rightarrow -\eta_t\,( {\cal Z}+{\cal Z}_b+h_c) $ and, analogously (not shown), $Y \rightarrow \theta_t\,( {\cal Z}+{\cal Z}_b+h_c) $.
All other transfer functions tend to zero at low frequency, 
 and therefore do not affect the observables in the range of interest.

{Using the  above described transfer functions, too cumbersome to be written here, it is possible to calculate the spectra of  the observables (in particular, $X$), as expected from tilt measurements. This allows us to assess the role of tilt  in the low-frequency  (below torsional resonances) noise,  
and its predicted effect on the $X$ variable. Tilt is indeed responsible for the observed high level of off-resonance noise in  the measured spectrum}

Incidentally, we note that the torsional observable $\varphi$ responds to tilt motion $\eta_t$ at the swinging modes ($\nu_3 \div \nu_6$): this is, probably, the reason why these resonances are observed in the  $\varphi$ data, while not predicted by the solution of free motion.

An analogous analysis can be carried on to evaluate the additional effect of translational floor motion.  This is not as interesting because these transfer functions would all vanish toward zero frequency, when the suspension point rigidly translates together with the entire enclosure and, in particular, with the GRS. Besides, as there is no effective way to measure such motion, calculation of  these  transfer functions is of little use.


\subsection{Low frequency behaviour: the 2 $soft$ DoF} \label{Lorenzo}
In many practical cases,  we are mainly interested in the  low frequencies dynamics of the system, i.e. in the region around its two $soft$, torsional resonances, namely between 1 and 3 mHz.
It can then  be useful to simplify our model in the limit where all swinging and bouncing motions can be neglected:  we are left with the two torsional motions described by the two angles $\varphi_a$ and $\varphi_b$, and governed by the equations:

\begin{eqnarray}
\ddot{\varphi }_a \left( I^b_{33}+d y_a \left(m_a+m_b\right)+I^a_{33}\right)+I^b_{33} \ddot{\varphi }_b+k_a \varphi_a&=&Q_{\varphi,a}\\
I^b_{33} \ddot{\varphi }_a+I^b_{33} \ddot{\varphi }_b+k_b \varphi_b&=&Q_{\varphi,b} \nonumber
\end{eqnarray}
with $Q_{\varphi_a}=M_{a,z}+d F_{b,x}+M_{b,z}$,\,and $Q_{\varphi_b}=M_{b,z}$.\\
It might be useful to re-express the equations in terms of the observables measured by the GRS (and defined in the Inertial System). In the limit considered, where the crossbar and TM move rigidly together (except for torsions),  we have \\
\begin{equation}
  X_{GRS} = \varphi_a  d \hskip 3cm \varphi = \varphi_a+\varphi_b
\end{equation}
The corresponding quadratic Lagrangian is
\be {\cal L}_T =  \frac12 I_c {\dot \varphi_a}^2 + \frac12 I_t {\dot \varphi}^2 - \frac12 k_a \varphi_a^2 - \frac12 k_b (\varphi_a - \varphi)^2, \ee
where  $I_t = I^b_{33}$ and $I_c=I^a_{33}+d y_a(m_a+m_b)$.

The (forced) Euler-Lagrange equations obtained from $ {\cal L}_T $ are
\ba
&&  I_c {\ddot \varphi_a} + k_a {\varphi_a} - k_b (\varphi - \varphi_a) = M_a,\label{L1}\\
&&  I_t {\ddot \varphi} +k_b (\varphi - \varphi_a) = M_b.\label{L2}
\ea
This simplified model is adequate to describe most of  behaviour of the double pendulum at frequencies near or below the torsion resonances.  However, its predictions might result inaccurate in instances where disturbances feed into the low frequency part of the spectrum through coupling with higher modes, like in the example of the previous section.

\section{Conclusions}

We have described an 8 DoF Lagrangian model that provides a suitable account for the motion of the double torsion pendulum ``PETER".
The model  fully describes its free dynamics and its response to external
disturbances, and can  accurately predict 
the torsional, swinging pendulum and bouncing
resonances. The model is otherwise `necessary' to
obtain a correct account of  external forces and torques acting on the system.
In particular, it gives a good account of the tilt noise, which feeds into the low
frequency motion of the double pendulum via a non-trivial
coupling among the various DoF. The number and location of resonance
peaks are correctly predicted and are used for a first validation of the model with preliminary
data. The transfer functions concerning seismic noise components
allow us to evaluate the response of the observables, with particular
emphasis on the `soft' translational motion. 
 Dissipation effects were not considered here, in order to keep the equations manageable.  However,  it is possible to extend the
Lagrangian formalism, via the Rayleigh dissipation function, to account for frictional forces. This is actually straightforward in the low-frequency, two DoF limit {(cfr. sect. \ref{Lorenzo})}, where dissipative effects are taken into account in order to predict, via the fluctuation-dissipation theorem, the fundamental limits of sensitivity of the system {\cite{parigi}}.

\appendix
\section{Equations of motion in Lagrangian coordinates.}
We define the following quantities, that are, in essence, modified moments of inertia (the first six are, e.g., referred to the upper suspension point).  
The last term is proportional to the $z$ coordinate of its center of mass.\\
\be \label{def_I}
\begin{array}{ll}
{\cal I}_{a,1}=I^a_{11}+m_a\left[ ({\cal Z}+{\cal Z}_a){}^2+y_a^2)\right], \\[8pt]  
{\cal I}_{b,1}=I^b_{11}+m_b\left[ ({\cal Z}+{\cal Z}_b){}^2+d^2)\right],\\[8pt]
{\cal I}_{a,2}=I^a_{22}+m_a ({\cal Z}+{\cal Z}_a){}^2, \\[8pt]  {\cal I}_{b,2}=I^b_{22}+m_b ({\cal Z}+{\cal Z}_b){}^2,\\[8pt]
{\cal I}_{a,3}=I^a_{33}+m_a y_a ^2, \\[8pt]   {\cal I}_{b,3}=I^b_{33}+m_b d^2, \\[8pt]  {\cal I}_{b,4}= I^b_{11}+m_b {\cal Z}_b^2,\\[8pt]
{\cal I}_{a,4}=I^a_{23}+m_b d \left({\cal Z}_a-{\cal Z}_b\right), \\[8pt]   {\cal I}_{b,5}= I^b_{11}+m_b {\cal Z}_b\left({\cal Z}+{\cal Z}_b\right),\\[8pt]
 c_1= m_b \left({\cal Z}+{\cal Z}_b\right)+m_a\left({\cal Z}+{\cal Z}_a\right).\\[8pt]
\end{array}
\ee

By assuming $I^b_{22}=I^b_{11};~I^a_{12}=I^a_{13}=0$, the equations of motion for the double pendulum are:
\be\label{EM}
\begin{array}{l}
({\cal I}_{a,1}+{\cal I}_{b,1})\ddot{\theta }_a +{\cal I}_{b,5} \ddot{\theta }_b+m_b d \ddot{\delta }_b + g c_1 \theta _a+m_b g {\cal Z}_b \theta _b=Q_{\theta ,a},\\[8pt]

 ({\cal I}_{a,2}+{\cal I}_{b,2})\ddot{\eta }_a+{\cal I}_{a,4}\ddot{\varphi }_a+{\cal I}_{b,5} \ddot{\eta }_b+ g c_1 \eta _a+m_b g{\cal Z}_b  \eta _b=Q_{\eta ,a},\\[8pt]

({\cal I}_{a,3}+{\cal I}_{b,3})\ddot{\varphi }_a +I^b_{33} \ddot{\varphi }_b+{\cal I}_{a,4}\ddot{\eta }_a -m_b d  {\cal Z}_b \ddot{\eta }_b+k_a \varphi _a=Q_{\varphi ,a},\\[8pt]

\left(m_a+m_b\right) \ddot{\delta }_a+m_b \ddot{\delta }_b+\delta _a \kappa_{e,a}=Q_{\delta ,a},\\  [8pt]

{\cal I}_{b,5} \ddot{\theta }_a + {\cal I}_{b,4}\ddot{\theta }_b+m_b g {\cal Z}_b (\theta _a +\theta _b)=Q_{\theta ,b},\\  [8pt]

{\cal I}_{b,5} \ddot{\eta }_a+{\cal I}_{b,4}\ddot{\eta }_b-m_b d {\cal Z}_b \ddot{\varphi }_a + m_b g {\cal Z}_b (\eta _a +\eta _b)=Q_{\eta,b},\\  [8pt]

I^b_{33} \ddot{\varphi }_a+I^b_{33} \ddot{\varphi }_b+k_b \varphi _b=Q_{\varphi ,b},\\[8pt]

m_b \ddot{\delta }_a+m_b\ddot{\delta }_b+d m_b \ddot{\theta }_a+\delta _b \kappa_{e,b}=Q_{\delta ,b}.\\[8pt]
\end{array}
\ee

Generalized forces  in Lagrangian coordinates are\\
 $
\begin{array}{l}
Q_{\theta_a}=\theta _a \left(- y_a F_{a,y}+\left({\cal Z}+{\cal Z}_b\right) F_{b,z}+\left({\cal Z}+{\cal Z}_a\right) F_{a,z}+d F_{b,y}\right)+\\
\qquad+y_a F_{a,z}+\delta _a \left(F_{a,y}+F_{b,y}\right)+\left({\cal Z}+{\cal Z}_b\right) F_{b,y}+\left({\cal Z}+{\cal Z}_a\right) F_{a,y}+\\
\qquad+M_{a,x}-d F_{b,z}+\delta _b F_{b,y}+\theta _b {\cal Z}_b F_{b,z}+M_{b,x},\\
\\[8pt]
Q_{\eta_a}=\dfrac{d \varphi _a m_b F_{a,z}}{m_a}-d \varphi _a F_{b,z}-\delta _a \left(F_{a,x}+F_{b,x}\right)-\left({\cal Z}+{\cal Z}_b\right) F_{b,x}+\\
\qquad+\eta _a \left(\left({\cal Z}+{\cal Z}_b\right)
   F_{b,z}+\left({\cal Z}+{\cal Z}_a\right) F_{a,z}\right)+\\
   \qquad+\theta _a \left(M_{a,z}+M_{b,z}\right)-\left({\cal Z}+{\cal Z}_a\right) F_{a,x}+M_{a,y}-\delta _b F_{b,x}+\eta _b {\cal Z}_b
   F_{b,z}+M_{b,y},\\
   \\[8pt]
      Q_{\varphi_a}= -\dfrac{d m_b \left(F_{a,x}+\varphi _a F_{a,y}-\eta _a F_{a,z}\right)}{m_a}+d \varphi _a F_{b,y}-d \eta _a F_{b,z}+\eta _a M_{b,x}-\\
   \qquad-\theta _a
   \left(M_{a,y}+M_{b,y}\right)+\eta _a M_{a,x}+M_{a,z}+F_{b,x} \left(d-\theta _b {\cal Z}_b\right)-\eta _b {\cal Z}_b F_{b,y}+M_{b,z},\\
   \\[8pt]
   Q_{\delta_a}=-\eta _a \left(F_{a,x}+F_{b,x}\right)+\theta _a \left(F_{a,y}+F_{b,y}\right)-F_{a,z}-F_{b,z},\\
   \\[8pt]
   Q_{\theta_b}={\cal Z}_b \left(-\varphi _a F_{b,x}+\left(\theta _a+\theta _b\right) F_{b,z}+F_{b,y}\right)+\varphi _a M_{b,y}-\eta _a M_{b,z}+\delta _b F_{b,y}+M_{b,x},\\
   \\[8pt]
   Q_{\eta_b}={\cal Z}_b \left(-\varphi _a F_{b,y}+\left(\eta _a+\eta _b\right) F_{b,z}-F_{b,x}\right)-\varphi _a M_{b,x}+\left(\theta _a+\theta _b\right) M_{b,z}-\delta _b
   F_{b,x}+M_{b,y},\\
   \\[8pt]
  Q_{\varphi_b}= \left(\eta _a+\eta _b\right) M_{b,x}-\left(\theta _a+\theta _b\right) M_{b,y}+M_{b,z},\\
   \\[8pt]
  Q_{\delta_b}=-\left(\eta _a+\eta _b\right) F_{b,x}+\left(\theta _a+\theta _b\right) F_{b,y}-F_{b,z}.\\
     \end{array}
$

We actually used a further simplified version of these expressions, where we only retain  terms of order  zero in the configuration variables:
\begin{eqnarray*}
Q_{\theta,a} &\simeq & F_{b,y} \left({\cal Z}+{\cal Z}_b\right)+F_{a,y} \left({\cal Z}+{\cal Z}_a\right)+M_{a,x}+M_{b,x}+y_a F_{a,z}-d F_{b,z},\\
Q_{\eta,a} &\simeq &-F_{b,x} \left({\cal Z}+{\cal Z}_b\right)-F_{a,x} \left({\cal Z}+{\cal Z}_a\right)+M_{a,y}+M_{b,y},\\
Q_{\varphi,a} &\simeq &- y_a F_{a,x}+d F_{b,x}+M_{a,z}+M_{b,z},\\
Q_{\delta,a} &\simeq &-F_{a,z}-F_{b,z},\\
Q_{\theta,b} &\simeq &F_{b,y} {\cal Z}_b+M_{b,x},\\
Q_{\eta,b} &\simeq &-F_{b,x} {\cal Z}_b+M_{b,y},\\
Q_{\varphi,b} &\simeq &M_{b,z},\\
Q_{\delta,b} &\simeq &-F_{b,z}.
\end{eqnarray*}



\section*{Acknowledgments}

The continued advice and support of S.Vitale, the Trento LISA-Pathfinder group, and, in particular, W. J. Weber   is appreciated. Work supported by INFN and by MIUR (grant PRIN 2008).

\bibliographystyle{amsplain}

\end{document}